\documentclass[rnote,traditabstract]{aa} 
\usepackage{graphicx}
\usepackage{txfonts}
\begin{document}
\title{Fermi/LAT detection of extraordinary variability in the gamma-ray emission of the blazar PKS 1510-089}

\author{L. Foschini\inst{1}, 
G. Bonnoli\inst{1}, 
G. Ghisellini\inst{1}, 
G. Tagliaferri\inst{1}, 
F. Tavecchio\inst{1}, 
A. Stamerra\inst{2,3}
}

\institute{
INAF - Osservatorio Astronomico di Brera, via E. Bianchi 46, 23807, Merate (LC), Italy\\
\email{luigi.foschini@brera.inaf.it}
\and
INAF - Osservatorio Astrofisico di Torino, via P. Giuria 1, 10125, Torino, Italy
\and
INFN - Sezione di Torino, via P. Giuria 1, 10125 Torino, Italy                     
             }

   \date{Received --; accepted --}
 
\abstract{We have reanalyzed the giant outburst of the blazar PKS~1510$-$089 ($z=0.36$) that occurred on 2011 October-November. The $\gamma$-ray flux in the $0.1-100$~GeV energy range exceeded the value of $10^{-5}$~ph~cm$^{-2}$~s$^{-1}$ for several days. The peak flux was reached on 2011 October 19, with a value of $\sim 4.4\times 10^{-5}$~ph~cm$^{-2}$~s$^{-1}$, which in turn corresponds to a luminosity of $\sim 2\times 10^{49}$~erg~s$^{-1}$. A very short timescale variability was measured. Particularly on 2011 October 18, the flux-doubling time was as short as $\sim 20$~minutes. This is the shortest variability ever detected in the MeV-GeV energy band. We compared our analysis with two other outbursts observed in 2009 March and 2012 February-March, when the blazar was also detected by HESS and MAGIC to infer information about the emission at hundreds of GeV.}

\keywords{galaxies: quasars: individual: PKS~B1510$-$089 -- galaxies: jets -- gamma-rays: galaxies}

\authorrunning{L. Foschini et al.}
\titlerunning{Extraordinary Variability in PKS 1510-089}

\maketitle

\section{Introduction}
According to current knowledge, the structure of relativistic jets in active galactic nuclei is self-similar (e.g. Heinz \& Sunyaev 2003). This means that the size of the emitting region $r$ is linked with the distance $R$ from the central supermassive black hole with the relationship $R \sim r/\psi$, where $\psi$ is a scaling factor generally with values between $0.1$ and $0.25$ (e.g. Ghisellini \& Tavecchio 2009, Dermer et al. 2009), although radio observations suggest even lower values (e.g. Jorstad et al. 2005, Pushkarev et al. 2009). Since it is expected that the central black hole is the source of perturbation along the jet, the minimum size of the blob has to be larger than the gravitational radius $r_{\rm g}=GM/c^2$, where $M$ is the mass of the compact object, $G$ the gravitational constant, and $c$ is the speed of light in vacuum. Given these constraints, the changes in the electromagnetic emission have to be on timescales $\tau/(1+z)>r/c\delta$, where $z$ is the redshift, and $\delta$ is the Doppler factor of the jet. For a typical blazar with black-hole mass of $10^{9}M_{\odot}$, Doppler factor $\delta \sim 10$, and dissipation region located at $R\sim 10^{3}r_{\rm g}$ (Ghisellini et al. 2010) -- which in turn means $r\sim 2\times 10^{16}$~cm with $\psi = 0.1$ -- the observed timescale $\tau$ has to be longer than $\sim 18$~hours.

Shorter timescales have previously been observed in the past with different facilities (Gaidos et al. 1996, Foschini et al. 2006, Aharonian et al. 2007, Albert et al. 2007). The launch of the {\it Fermi Gamma-ray Space Telescope} (hereafter {\it Fermi}) in 2008 made it possible to significantly increase the sample of observed blazars that display intraday variability in the $\gamma$-ray band, reaching values shorter than $2-3$ hours (e.g. Tavecchio et al. 2010, Foschini et al. 2011a,b, Vovk \& Neronov 2013). The flat-spectrum radio quasar (FSRQ) PKS~1222$+$216 ($z=0.432$) is particularly interesting with its minimum observed timescale of $1.0\pm 0.2$~hours ($\sim 0.7$~hours intrinsic) measured by {\it Fermi} during the outburst of 2010 April 30 (Foschini et al. 2011b). After about 1.5 months, on 2010 June 17, the MAGIC \v Cerenkov telescope detected the same source at hundreds of GeV with a variability timescale of 10 minutes (Aleksi\'c et al. 2011). The two outbursts were also marked by a spectral change: during the April outburst, the $\gamma$-ray spectrum was soft, with a cut-off and the energy of the highest detected photon equal to $\sim 23$~GeV; in the second outburst in June the spectrum was hard and extended up to a few hundreds of GeV. This has been interpreted as a change of the dissipation location: within the broad-line region (BLR) during the April outburst and outside it during the June one (Foschini et al. 2011b; but see Tavecchio et al. 2011 for another interpretation). 

Recently, Ghisellini et al. (2013) reported about PMN~J2345$-$1555 ($z=0.621$): for the first time, the multiwavelength coverage allowed us to understand that the hardening of the $\gamma$-ray spectrum was linked to a shift of the synchrotron peak at higher energies in the X-rays, confirming the change of its nature from red to blue (see Ghisellini \& Tavecchio 2008 for an explanation of blazar colors). Other hints of changes in the blazar color were reported for PKS~2155$-$304 ($z=0.116$) in 2006 November (Foschini et al. 2008, but no $\gamma$-ray data were available) and PKS~1510$-$089 ($z=0.36$) in 2009 March (Abdo et al. 2010, D'Ammando et al. 2011). In the latter case, the peak of the synchrotron shifted from infrared to optical/UV frequencies and the source was detected at hundreds of GeV by the HESS Collaboration (2013), although the X-ray spectrum did not change (Abdo et al. 2010). 

\begin{figure*}[t]
\centering
\includegraphics[angle=270,scale=0.70]{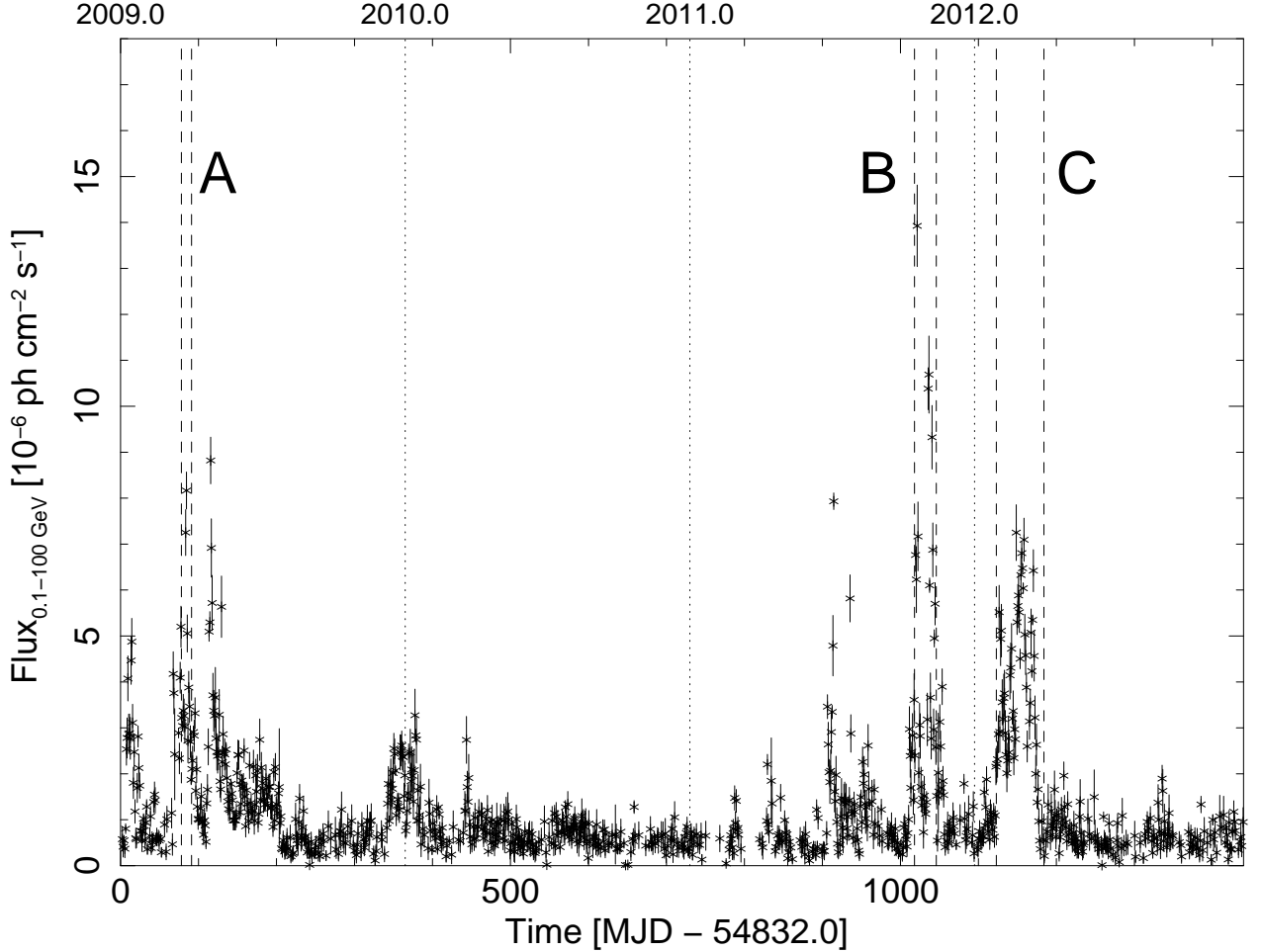}
\caption{Global light curve of PKS~1510$-$089 in the $0.1<E<100$~GeV energy band with 1-day time bin. Time starts on 2009 January 1 (MJD 54832.0). Vertical dotted lines indicated the 1 January of each year. Vertical dashed lines delimit the period of the three outbursts analyzed here. Outburst A: 2009 March; outburst B: 2011 October/November; outburst C: 2012 February/March.}
\label{fig:global}
\end{figure*}

Yet PKS~1510$-$089 displayed very interesting activity in 2011 October-November with $\gamma$-ray flux measured by {\it Fermi} exceeding $10^{-5}$~ph~cm$^{-2}$~s$^{-1}$ and the shorter timescale of variability of about one hour (Brown 2013, Saito et al. 2013). There was no multiwavelength coverage, because the source was apparently too close to the Sun, except for radio observations reporting an extraordinary increase of the flux density at different frequencies (Nestoras et al. 2011, Orienti et al. 2011, Beaklini et al. 2011). Nevertheless, it is possible to draw useful inferences from the study of {\it Fermi} LAT data alone. Here we report a reanalysis of the $\gamma$-ray emission in the same period, but with the addition of a different method to extract the light curves with time bins shorter than three hours. This allowed us to investigate time variability on scales shorter than those reported by Brown (2013) and Saito et al. (2013). A comparison with other outbursts (2009 March, 2012 February/March), moreover, suggests some interesting features of the jet of PKS~1510$-$089. 

PKS~1510$-$089 has redshift $z=0.36$ (Thompson et al. 1990). In the usual $\Lambda$CDM cosmology with the latest measured values of the Hubble-Lema\^itre constant $H_{0}=70$~km~s$^{-1}$~Mpc$^{-1}$ and $\Omega_{\Lambda}=0.73$ (Komatsu et al. 2011), the luminosity distance is 1934~Mpc, and 1 arcsec corresponds to 5.1~kpc.

\section{Data analysis and discussion}
Fig.~\ref{fig:global} shows the $\gamma$-ray emission as measured by {\it Fermi}/LAT of PKS~1510$-$089 over a period of about four years. In the present work, we focused mostly on outburst B (2011 October/November), but we compared it also with outbursts A and C (2009 March and 2012 February/March, respectively) to draw useful inferences on the source properties.

\begin{figure}[th]
\centering
\includegraphics[angle=270,scale=0.32]{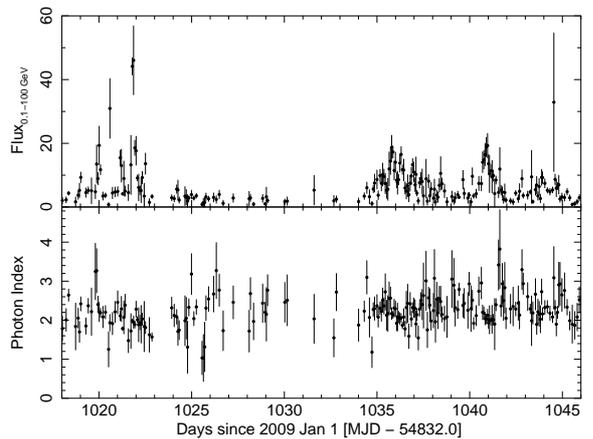}
\caption{Outburst B (2011 October/November): Light curves of PKS~1510$-$089 in the $0.1<E<100$~GeV energy band with GTI-bin. (\emph{top panel}) Flux in units of [$10^{-6}$~ph~cm$^{-2}$~s$^{-1}$]. (\emph{bottom panel}) Photon index.}
\label{fig:2011}
\end{figure}

The analysis of the \emph{Fermi}/LAT data was performed following the standard procedures\footnote{E.g. http://fermi.gsfc.nasa.gov/ssc/data/analysis/scitools/} for light curves with time bins longer than three hours (i.e., two orbits), which is the minimum integration time for a complete coverage of the whole sky (Fig.~\ref{fig:global}). In addition, we also built light curves with shorter time bins (e.g., Fig.~\ref{fig:2011}) using the method described in Foschini et al. (2011a,b). In that case, the time bin is equal to one good-time interval (GTI), which in turn is shorter than one orbit ($\sim 95$~min) and can have variable duration depending on the pointing direction. In both cases, the flux was estimated by using the {\tt gtlike} task of the {\tt LAT Science Tools}. The methods were applied by using the most recent version of the software and calibration: {\tt LAT Science Tools v. 9.27.1}, instrument response function (IRF) {\tt P7SOURCE\_V6}, isotropic background {\tt isop7v6source.txt}, and Galactic diffuse background {\tt gal2yearp7v6v0.fits}. PKS~1510$-$089 and the contaminating sources within $10^{\circ}$ from its sky position were modeled by using a simple power law $F(E)\propto E^{-\Gamma}$, where $\Gamma$ is the photon index. The threshold of Test Statistic (TS, see Mattox et al. 1997) for a meaningful detection was set to 9 ($\sim 3\sigma$). 

The peak flux was measured on 2011 October 19 19:06 UTC (MJD$=55853.79595$, source ontime~$\sim 2.2$~ks) as $(4.4\pm 0.3)\times 10^{-5}$~ph~cm$^{-2}$~s$^{-1}$, with $\Gamma = 1.98\pm 0.04$ ($TS=598$) and 119 counts recorded. This corresponds to a luminosity of about $2\times 10^{49}$~erg~s$^{-1}$ making PKS~1510$-$089 one of the brightest blazars ever detected at $\gamma$ rays (the record remains to 3C~454.3, with $3\times 10^{50}$~erg~s$^{-1}$ on 2010 November 20; see Foschini et al. 2011a).

The light curves were scanned to search for the minimum time of doubling/halving flux between two consecutive points,

\begin{equation}
\frac{F(t_1)}{F(t_0)} = 2^{-(t_1-t_0)/\tau},
\end{equation}

\noindent where $F(t_0)$ and $F(t_1)$ are the fluxes at the two consecutive times $t_0$ and $t_1$, respectively, and $\tau$ is the extrapolated doubling/halving timescale. To obtain a meaningful measurement, we set a $3\sigma$ threshold in the flux change between the two consecutive measurements. The results are displayed in Table~\ref{tab:timescales} and are limited to the lowest values of about one hour or shorter. Table~\ref{tab:timescales} also reports the values of $\tau$ as measured from the light curve with a three-hour time bin (not shown) to compare and confirm the results published by Brown (2013) and Saito et al. (2013). 
We noted that the shortest $\tau$ are for increasing flux, while the lowest value for the decay time is an upper limit of five hours. This is consistent with the results of Nalewajko (2013), who found that PKS~1510$-$089 has the most skewed flares, with short rise-times and long decays. 

\begin{table*}
\caption{Summary of doubling/halving times measured. Dates are given in YYYY/MM/DD; $t_0$ and $t_1$ are in [MJD], while half-bin times are in hours; fluxes are in units of [$10^{-6}$~ph~cm$^{-2}$~s$^{-1}$] and the TS of the detection is indicated between parentheses; the significance S of the flux difference is in [$\sigma$]; the absolute values of the observed characteristic timescale $\tau$ (positive for decay, negative for rise) and the intrinsic one $\tau_{\rm int}=\tau/(1+z)$ are given in hours.}
\begin{center}
\begin{tabular}{cccccccccc}
\hline
date & $t_0$ & bin & $t_1$ & bin & $F(t_0)$ & $F(t_1)$ & S & $\tau$ & $\tau_{\rm int}$\\
\hline
\multicolumn{10}{c}{{\it Outburst A (2009) -- GTI time bin}}\\
\hline
2009/03/31 & 54921.18046 & 0.79 & 54921.24677 & 0.79 & $2.44\pm 0.53$ (34) & $7.88\pm 1.04$ (18) & 5.3 & $-1.36\pm 0.71$ & $-1.00\pm 0.52$\\
\hline
\multicolumn{10}{c}{{\it Outburst B (2011) -- 3-hr time bin}}\\
\hline
2011/10/18 & 55852.43752 & 1.5 & 55852.56252 & 1.5 & $0.66\pm 0.58$ (22) & $11.19\pm 3.10$ (60) & 3.4 & $-1.06\pm 0.09$ & $-0.78\pm0.07$ \\ 
2011/10/19 & 55853.68752 & 1.5 & 55853.81252 & 1.5 & $8.52\pm 5.95$ (37) & $44.30\pm 4.73$ (609) & 6.0 & $-1.82\pm 0.21$ & $-1.34\pm 0.15$ \\
\hline
\multicolumn{10}{c}{{\it Outburst B (2011) -- GTI time bin}}\\
\hline
2011/10/18 & 55852.52144 & 0.65 & 55852.59130 & 0.61 & $0.75\pm 0.66$ (22) & $30.93\pm 9.33$ (95) & 3.2 & $-0.45\pm 0.03$ & $-0.33\pm 0.02$\\
2011/11/01 & 55866.73018 & 0.67 & 55866.79917 & 0.79 & $1.06\pm 0.75$ (38) & $5.67\pm 1.18$ (35) & 3.9 & $-0.99\pm 0.02$ & $-0.73\pm 0.01$\\
2011/11/04 & 55869.71600 & 0.77 & 55869.78141 & 0.79 & $1.85\pm 1.25$ (15) & $8.12\pm 2.06$ (51) & 3.0 & $-1.06\pm 0.05$ & $-0.78\pm 0.04$\\
\hline
\multicolumn{10}{c}{{\it Outburst C (2012) -- GTI time bin}}\\
\hline
2012/03/03 & 55989.90896 & 0.63 & 55989.97872 & 0.65 & $0.76\pm 0.47$ (12) & $4.88\pm 0.49$ (72) & 8.5 & $-0.90\pm 0.11$ & $-0.66\pm 0.08$\\
\hline
\end{tabular}
\end{center}
\label{tab:timescales}
\end{table*}

\begin{figure}[th]
\centering
\includegraphics[angle=270,scale=0.32]{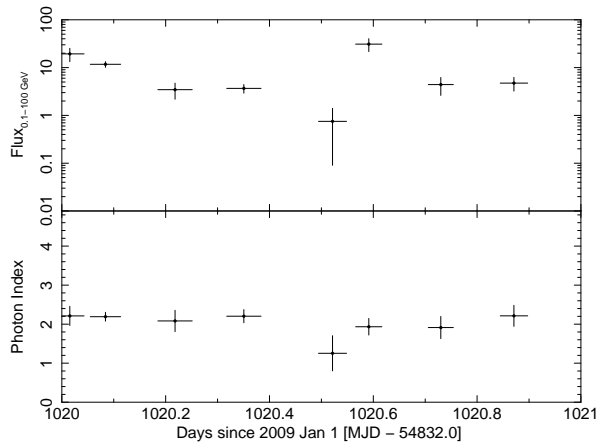}
\caption{Zoom of outburst B (2011 October/November): Light curves of PKS~1510$-$089 in the $0.1<E<100$~GeV energy band with GTI-bin centered on the shortest timescale flare. (\emph{top panel}) Flux in units of [$10^{-6}$~ph~cm$^{-2}$~s$^{-1}$] in log scale to emphasize the variability. (\emph{bottom panel}) Photon index.}
\label{fig:2011zoom}
\end{figure}

The shortest doubling time is $\sim 20$~minutes ($\tau_{\rm int}=20\pm1$~min), measured during the outburst B (Fig.~\ref{fig:2011zoom}) between MJD$=55852.52144$ (2011 October 18 12:30 UTC) and MJD$=55852.59130$ (2011 October 18 14:11 UTC). This corresponds to a size of the emitting region equal to $r\sim 4\times 10^{14}$~cm (by assuming $\delta=10$). This value has to be compared with the gravitational radius of PKS~1510$-089$, which is $r_{\rm g}\sim 3\times 10^{13}$~cm (by assuming a mass of the central black hole of $2\times 10^{8}M_{\odot}$, from Xie et al. 2005). 

During that episode, PKS~1510$-$089 jumped by almost two orders of magnitude in less than one orbit from a flux of $(0.75\pm 0.66)\times 10^{-6}$~ph~cm$^{-2}$~s$^{-1}$ ($TS=22$) to $(31\pm 9)\times 10^{-6}$~ph~cm$^{-2}$~s$^{-1}$ ($TS=95$). Interestingly, the photon index at low flux was hard ($\Gamma=1.2\pm 0.4$) and became softer at high flux ($\Gamma=1.9\pm 0.2$), although the highest detected photon energy was $\sim 14$~GeV. No photon above 100~GeV was detected during the entire studied period of outburst B to confirm a really hard spectrum extending to hundreds of GeV or with a cut-off at tens of GeV, as expected if the dissipation occurs inside the broad-line region. Neither \v Cerenkov observation nor X-ray follow-up (to search for a synchrotron peak shift) was possible at that epoch, because of the Sun constraints. 


To better assess if the lack of very high energy photons could be due to the very short exposures and small LAT effective area above 100 GeV, we compared the event with the 2009 March outburst (outburst A). This event was previously described in other works (e.g. Abdo et al. 2010, Marscher et al. 2010, Tavecchio et al. 2010). The authors reported continuously a very soft photon index of the $\gamma$-ray spectrum ($\Gamma>2$). Abdo et al. (2010) reported a hardening trend of the photon index with increasing flux, but still from $\sim 2.6$ to $\sim 2.2$. Therefore, comparing this information with our analysis, it seems that outburst A (2009) was different from outburst B (2011). However, in 2009 March (A), the blazar was detected at hundreds of GeV by the HESS Collaboration (2013), thus requiring a hard photon index at MeV-GeV energies. Therefore, we reanalyzed the LAT data starting from 2009 March 20 (MJD 54910.0) and ending after a couple of weeks by using the GTI-bin light curve (not shown). We noted that the shorter doubling time is about one hour and the photon index now sometimes became harder ($\Gamma < 2$), as expected.

The blazar was also detected at hundreds of GeV also by MAGIC during outburst C (2012 February/March, Lindfors et al. 2013). Therefore, we also reanalyzed the LAT data for the period from 2012 January 29 00:00 UTC (MJD 55955.0) to April 3 00:00 UTC (MJD 56020.0) by using the GTI-bin light curve (not shown). The shorter timescale is about one hour (Table~\ref{tab:timescales}) and was measured while changing from a low hard ($\Gamma = 1.5\pm 0.3$) to high soft ($\Gamma=2.2\pm 0.1$) flux, just as happened in 2011. It is worth noting that this event occurred during one successful exposure with MAGIC (cf Fig.~5 of Lindfors et al. 2013).

\begin{figure}[th!]
\centering
\includegraphics[angle=270,scale=0.32]{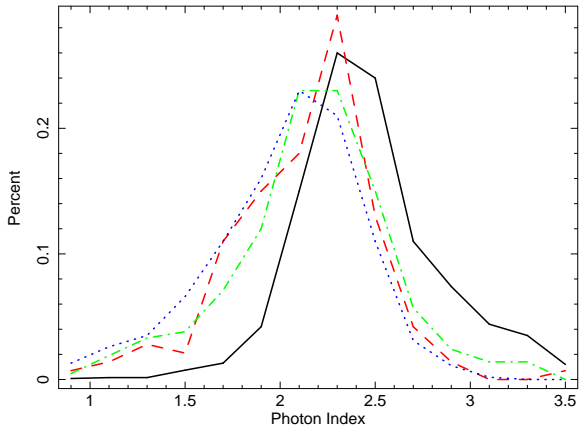}
\caption{Distribution of the photon indexes as measured by using the one-day light curve over four years of activity (black continuum line), and the GTI-bin light curves for the outbursts A (red dashed line), B (green dot-dashed line), and C (blue dotted line). It is evident that the distributions during the three outbursts are harder.}
\label{fig:photonindex}
\end{figure}

It is worth noting that the hard photon indexes measured during these episodes of short timescale variability are not isolated. Fig.~\ref{fig:photonindex} shows the distributions of the photon indexes during the three outbursts (measured from the GTI-bin light curves) compared with the distribution of $\Gamma$ measured on the one-day bin light curve shown in Fig.~\ref{fig:global}. The drift toward harder indexes during the outbursts and as measured on short timescales is evident.

Therefore, given the similarities of the three outbursts, we conclude that: (i) it is reasonable to assume that PKS~1510$-$089 could have been detected at hundreds of GeV also in 2011 (i.e., the photon index as measured from the LAT data analysis are really hard and the lack of very high energy photons is due to the short exposures); (ii) the $\gamma$-ray spectrum can be hard on very short time-periods (less than three hours), while when integrated on longer exposures, the soft component dominates.

\section{Final remarks} 
The present giant outburst of PKS~1510$-$089 seems to be different from that of PKS~1222$+$216 (cf. Foschini et al. 2011b). The latter displayed two giant outbursts in 2010: the first one with a cut-off at tens of GeV and doubling flux timescale of about one hour; the second one extending without changes in the spectral slope up to hundreds of GeV and with variability of $\sim 10$~minutes as detected by MAGIC (Aleksi\'c et al. 2011). The two outbursts were separated by about 1.5~months, more than sufficient to assume a drift of the dissipation region from inside to outside the BLR. 

Instead, the 2011 outburst (B) of PKS~1510$-$089 reported here (see also Brown 2013, Saito et al. 2013) displayed different properties. The $0.1-100$~GeV flux well exceeded the value of $10^{-5}$~ph~cm$^{-2}$~s$^{-1}$ three times during the period of one month, with very short doubling times recorded in the first two bursts and hard photon indexes (Fig.~\ref{fig:2011}). Particularly the first two episodes, which were separated by about two weeks (hence sufficient to assume some significant drift of the dissipation region), displayed similar spectral and variability characteristics. Therefore, it is not possible to invoke an explanation similar to PKS~1222$+$216. To explain the hard photon indexes (and the detection at hundreds of GeV by \v Cerenkov telescopes) on short timescales, either a stable multi-zone structure, where the outbursts are due to disturbances moving down the jet (Marscher et al. 2010), or a structured jet, with compact blobs responsible of the shortest and most powerful flares (Tavecchio et al. 2011), are a more plausible explanations. Although a doubling time of $\sim 20$~minutes coupled with detections at hundreds of GeV remain a severe challenge for the current emission models of FSRQs (see discussions in Tavecchio et al. 2010, 2011; Foschini et al. 2011a).

Despite the limitations of the present work due to the lack of MW coverage, one firm conclusion can be, and has to be, stressed. Compared with other similar cases, this episode shows that there could be significant differences from source to source, and even from event to event of the same source. Therefore, individual case studies are still a powerful source of information to understand the nature of relativistic jets. It is of great importance to report as many as possible of these events to set up a significant archive. In this scenario, the continuous monitoring of the $\gamma$-ray sky performed by {\it Fermi} is a priceless font of data.

\section*{Acknowledgments}
This research has made use of publicly available {\it Fermi}/LAT data and analysis software obtained from the Fermi Science Support Center at the High Energy Astrophysics Science Archive Research Center (HEASARC), provided by NASA's Goddard Space Flight Center.

\end{document}